---------------------------------------------------------------------------
############################## .TEX FILE ###################################
---------------------------------------------------------------------------

\documentstyle[aps]{revtex}            
\begin{document}


\draft
\preprint{}
\title{Complex Geometry and Dirac Equation}
\author{Stefano De Leo\thanks{{\em E-Mail}: 
              {\tt deleos@le.infn.it} ,    
              {\tt deleo@ime.unicamp.br}    }$^{a,b}$,
        Waldyr A.~Rodrigues, Jr.\thanks{{\em E-Mail}:
              {\tt walrod@ime.unicamp.br}    }$^{b}$ and
        Jayme Vaz, Jr.\thanks{{\em E-Mail}:
              {\tt vaz@ime.unicamp.br}    }$^{b}$}
\address{$^{a}$Dipartimento di Fisica, Universit\`a degli Studi Lecce,\\
               Istituto Nazionale di Fisica Nucleare, INFN, Sezione di Lecce\\
                via Arnesano, CP 193, 73100 Lecce, Italia\\
                and\\ 
         $^{b}$Instituto de Matem\'atica, Estat\'{\i}stica e 
                Computa\c{c}\~ao Cient\'{\i}fica, IMECC-UNICAMP\\
                CP 6065, 13081-970, Campinas, S.P., Brasil} 
\date{\today}
\maketitle


\begin{abstract}

Complex geometry represents a fundamental ingredient in the formulation of 
the Dirac equation by the Clifford algebra. The choice of appropriate complex 
geometries is strictly related to the 
geometric interpretation of the complex imaginary unit $i=\sqrt{-1}$. 
We discuss {\em two} possibilities which appear in the multivector 
algebra approach: the $\sigma_{123}$ and $\sigma_{21}$ complex geometries.
Our formalism permits to perform a set of rules which allows an 
immediate translation between the complex standard Dirac theory and its  
version within geometric algebra. The problem concerning a double geometric 
interpretation for the complex imaginary unit $i=\sqrt{-1}$ is also discussed.


\end{abstract}


\section{Introduction}

In this paper we present a set of rules for passing back and forth between the
standard (complex) matrix-based approach to spinors in 4 dimensions and the 
geometric algebra formalism. This ``translation'' is only partial, consistent
with the fact that the Hestenes formalism~\cite{HES1} 
provides additional geometrical 
interpretations. In a pure translation nothing can be predicted which is 
not already in the original theory. In the new version of Dirac's equation 
some assumptions appear more natural, some calculations more rapid and new 
geometric interpretations for the complex imaginary unit $i=\sqrt{-1}$ 
appear in the 
translated version for the first time. 

The matrix form of spinor calculus and the vector calculus formulated by 
Gibbs can be replaced by a single mathematical system,
called multivector algebra, with which the tasks of theoretical physics can be
carried out more efficiently. The multivector algebra derives its power from 
the
fact that both the elements and the operations of the algebra are subject to
direct geometric interpretation~\cite{HH}. 
The geometric algebra is surely the most 
powerful and general language available for the development of mathematical
physics~\cite{HES2,LOUN}. 
The central result is a representation of the Dirac wave function 
which reveals a geometric structure, hidden in the conventional 
formulation~\cite{HES3}.

``The projection of the Dirac equation into the Pauli algebra eliminates 
redundancies, simplifying our task to solve this equation, since in the Pauli 
algebra we work in an eight dimensional space over the real numbers, while 
in the standard formulation we have to do with a 32-dimensional space over the
reals, the space of $4\times 4$ complex matrix ${\cal C}_{(4)}$''. 
- {\em Zeni}~\cite{ZEN}.

``The imaginary unit appearing in the Dirac equation and the energy-momentum 
operator represents the bivector generator of rotations in a space-like 
plane corresponding to the direction of the electron spin''. 
- {\em Hestenes}~\cite{HES4}.

We wish to clarify these statements. We agree with fact that 
in the Pauli algebra (isomorphic to the even part of the space/time algebra 
$Cl_{1,3}^+$) we have only 8
real parameters in defining the Dirac spinors, but in defining the most 
general operator which acts on them, how many real parameters do we need?
The imaginary unit $i$ is identified by the bivector 
$\sigma_{21} \in Cl_{3,0}$. Is this the
only opportunity? What about the possibility to identify the complex imaginary 
unit by the pseudoscalar $\sigma_{123} \in Cl_{3,0}$?

In formulating the Dirac equation by the Pauli algebra we can start from 
the standard matrix formulation and use the ideal approach to spinors 
to make a clear 
translation to the Clifford algebra $Cl_{4,1}$ which is isomorphic to 
$M_4({\cal C})$. The following step is to 
reduce the formulation of the Dirac equation to an algebra
of smaller dimension, the space-time algebra, $Cl_{1,3}$. Finally, we get a 
projection of the Dirac equation in the Pauli algebra $Cl_{3,0}$~\cite{ZEN}. 

In this paper
we shall perform a different approach. We give a set of rules which allow to
immediately write the Dirac equation by using the Pauli algebra. 
The fundamental ingredients of this translation are 
the direct identification of the complex imaginary unit $i=\sqrt{-1}$ 
by elements of 
the Pauli algebra and the introduction of the concept of ``complex'' 
geometry~\cite{REM,HB}.

The standard (complex) 4-dimensional spinor
\begin{equation}
\Psi \equiv 
\left( \begin{array}{c} \psi_1 \\ \psi_2 \\ \psi_3 \\ \psi_4
\end{array} \right) \equiv
\left( \begin{array}{c}  \varphi_1 + i \eta_1 \\
                         \varphi_2 + i \eta_2 \\
                         \varphi_3 + i \eta_3 \\
                         \varphi_4 + i \eta_4 
\end{array} \right)~~~~~~~\varphi_m , \eta_m \in {\cal R},~~~m=1,2,3,4~,
\end{equation}
is characterized by 8 real parameters, which can be settled
in the following 8-dimensional Clifford algebras
\[ Cl_{3,0}~[\sim M_2({\cal C})]~,~~~
   Cl_{1,2}~[\sim M_2({\cal C})]~,~~~
   Cl_{0,3}~[\sim {\cal H} \oplus {\cal H}]~,~~~
   Cl_{2,1}~[\sim M_2({\cal R}) \oplus M_2({\cal R})]~.
\]
The natural choice is $Cl_{3,0}~[\sim M_2({\cal C})]$, the 
algebra of the three-dimensional space. Such algebra allows an immediate 
geometric interpretation  for the Pauli matrices:
\begin{center}
\begin{tabular}{||lc||}\hline \hline
 &    ~~~~~$Cl_{3,0}$~~~~~\\
\hline
~scalar  
   & 1\\
~vectors  
& 
  ~~~$\sigma_1,~\sigma_2,~\sigma_3$~~~\\
~bivectors~~~ 
& 
  ~~$\sigma_2 \sigma_1 ,~\sigma_2 \sigma_3,~ \sigma_3 \sigma_1$~~ \\
~trivector   
& $\sigma_1 \sigma_2 \sigma_3$ \\
\hline \hline
\end{tabular}
\end{center}
The Pauli algebra can be also represented by the complexified quaternionic 
ring~\cite{DR,DR2}:
\begin{center}
\begin{tabular}{||c||}\hline \hline
 ~~~~~${\cal H}_c$~~~~~\\
\hline
 1 \\ 
~~~$\bbox{\iota} {\cal I},~\bbox{\iota} {\cal J} ,~\bbox{\iota} {\cal K}$~~~\\
 ${\cal I},~{\cal J},~{\cal K}$\\
$\bbox{\iota}$\\
\hline \hline
\end{tabular}
\end{center}
In the following, we prefer to use the vectors $\vec{\sigma} \in Cl_{3,0}$, 
in order to avoid confusion in the identification of the standard
(complex) imaginary unit $i=\sqrt{-1}$ by elements of the Pauli algebra.  
By identifying the complex imaginary unit $i=\sqrt{-1}$ by elements of 
$Cl_{3,0}$, we must recognize {\em two} possibilities 
\[ 
i=\sqrt{-1} ~~ \rightarrow ~~~~ 
\sigma_{21} \equiv \sigma_2 \sigma_1~\mbox{(bivector)~}
~~~~\mbox{or}~~~~~
\sigma_{123} \equiv \sigma_1 \sigma_2 \sigma_3 ~\mbox{(volume element)}~,
\]
in fact 
\[ \sigma_{21}^2=\sigma_{123}^2=-1~. \]
Consequently,  $\varphi_m + i \eta_m$ can be respectively translated by
\[
\varphi_m + \sigma_{21} \eta_m ~~~~~\mbox{or}~~~~~
\varphi_m + \sigma_{123} \eta_m~~~~~~~m=1,...,4~.
\]

We propose in this paper a discussion concerning these two different 
possibilities of translation for the standard complex Dirac theory. These two
possibilities are strictly related to the use of {\em two} different 
``complex'' geometries, namely
\begin{center}
the~~~$\sigma_{123}$~~~and~~~$\sigma_{21}$~~~complex geometries~.
\end{center}
In our formalism the standard physical results are soon reproduced. The 
possibility of choosing {\em two} different ``complex'' geometries in 
performing our translations will give an embarrassing situation: 
{\em two different geometric interpretations} for the complex imaginary 
unit $i=\sqrt{-1}$, namely
\begin{center}
bivector~~~or~~~volume element~.
\end{center}

\section{Probability Amplitudes and Complex Geometry}

The noncommutativity of the element of $Cl_{3,0}$ algebra requires to 
specify whether our Hilbert space, $V_{Cl_{3,0}}$, is 
to be performed by right or left 
multiplication of vectors by scalars. We will follow the usual choice
and work with a linear vector space under right multiplication by 
scalars~\cite{DR,FIN,ADL,HES5,LOU,KEL,GLD}. 
In quantum mechanics, probability amplitudes, rather than probabilities, 
superimpose, so we must determine what kinds of number system can be used 
for the probability amplitudes ${\cal A}$. We need a real modulus function
$N({\cal A})$ such that
\[ \mbox{Probability} = [N({\cal A})]^2~.\]
The first four assumptions on the modulus function are basically technical in
nature
\begin{eqnarray*}
N(0) = 0~,\\
N({\cal A})> 0 ~\mbox{if}~ {\cal A} \neq 0~,\\
N(r{\cal A}) = |r| N ({\cal A})~,~r~\mbox{real}~,\\
N({\cal A}_1 + {\cal A}_2) \leq N({\cal A}_1) + N({\cal A}_2)~.
\end{eqnarray*}
A final assumption about $N({\cal A})$ is physically motived by imposing the
{\em correspondence principle} in the following form: We require that in the 
absence of quantum interference effects, probability amplitude super-imposition
should reduce to probability super-imposition. So we have an additional 
condition on $N({\cal A})$:
\[ N({\cal A}_1 {\cal A}_2) =  N({\cal A}_1) N({\cal A}_2)~.\]
A remarkable theorem of Albert shows that the only algebras over the reals, 
admitting a modulus functions with the previous properties  are the reals
${\cal R}$, the complex ${\cal C}$, the (real) quaternions ${\cal H}$ and the 
octonions ${\cal O}$. The previous properties of the modulus function seem to
constrain us to work with {\em division algebras}  (which are finite 
dimensional algebras for which $a\neq 0$, $b\neq 0$ imply $ab \neq 0$), in 
fact
\[ {\cal A}_1 \neq 0~,~ {\cal A}_2\neq 0 \]
implies
\[ N({\cal A}_1 {\cal A}_2) = N({\cal A}_1) N({\cal A}_2) \neq 0 \]
which gives
\[ {\cal A}_1 {\cal A}_2 \neq 0~. \]
A simple example of non-division algebra is provided by the algebra 
$Cl_{3,0}$ since
\[ \left( 1 + \sigma_{3} \right)
   \left( 1 - \sigma_{3} \right) = 0
\]
guarantees that there are nonzero divisors of zero. So, if the probability
amplitudes are assumed to be element of $Cl_{3,0}$, we cannot give a 
satisfactory  probability interpretation. Nevertheless, we know that 
probability amplitudes are connected to inner products, thus, we can overcome
the above difficulty by defining an {\em appropriate} scalar product.

We have four possibilities:\\ 
We can define a binary mapping 
$\langle \Psi \mid \Phi \rangle$ of $V_{Cl_{3,0}} \times V_{Cl_{3,0}}$
into the scalar(S)/bivectorial(BV) part of $Cl_{3,0}$, 
we recall that $V_{Cl_{3,0}}$ 
represents the Hilbert space with elements defined in the Pauli algebra, 
\[ 
\langle \Psi \mid \Phi \rangle_{(S,BV)} = 
\left( \int \, d^3x \, \Psi^{\dag} \Phi \right)_{(S,BV)}~.
\] 
Note that the algebra $(1,\sigma_{21},\sigma_{23}, \sigma_{31})$ is isomorphic
to the quaternionic algebra. Thus, we have the mapping 
\[ V_{Cl_{3,0}} \times V_{Cl_{3,0}} ~~\rightarrow ~~ Cl_{0,2}~\sim {\cal H}~.\]
We can also adopt the more restrictive
``scalar'' projection $\langle \Psi \mid \Phi \rangle_{S}$:
\[ V_{Cl_{3,0}} \times V_{Cl_{3,0}} ~~\rightarrow ~~ 
Cl_{0,0}~\sim ~ {\cal R}~.\]
The last two 
possibilities are represented by the so-called ``complex'' geometries
\[ 
\langle \Psi \mid \Phi \rangle_{(1,\sigma_{21})}~~~~~\mbox{and}~~~~~
\langle \Psi \mid \Phi \rangle_{(1,\sigma_{123})}~.
\]
In these case we define the following binary mappings
\begin{eqnarray*} 
V_{Cl_{3,0}} \times V_{Cl_{3,0}} & ~~\rightarrow~~ &  
Cl_{0,1}^{i \rightarrow \sigma_{21}}~\equiv ~ {\cal C}(1,\sigma_{21})~,\\
V_{Cl_{3,0}} \times V_{Cl_{3,0}} & \rightarrow &  
Cl_{0,1}^{i \rightarrow \sigma_{123}} ~\equiv ~ {\cal C}(1,\sigma_{123})~.
\end{eqnarray*}
In the standard definition of inner product we find the operation of 
transpose conjugation, $\Psi^{\dag}$. How can we translate the transpose
conjugation in the geometric algebra formalism?

The Clifford algebra $Cl_{3,0}$ has three involutions similar to complex
conjugation. Take an arbitrary element
\[ E = E_0 + E_1 + E_2 + E_3 ~~~\mbox{in}~~~Cl_{3,0}~,\]
written as a sum of a scalar $E_0$, a vector $E_1$, a bivector $E_2$ and
a volume element $E_3$. We introduce the following involutions
\begin{eqnarray*}
E^{\bullet} & ~=~ & E_0 - E_1 + E_2 - E_3 ~~~~~~~\mbox{grade involution}~,\\
E^{\star}   & ~=~ & E_0 - E_1 - E_2 + E_3 ~~~~~~~\mbox{conjugation}~,\\
E^{\dag}    & ~=~ & E_0 + E_1 - E_2 - E_3 ~~~~~~~\mbox{reversion}~.
\end{eqnarray*}
The grade involution is an automorphism
\[ \left( E_a E_b \right)^{\bullet} = 
   E_a^{\bullet} E_b^{\bullet}~,
\]
while the reversion and the conjugation are anti-automorphism, that is,
\begin{eqnarray*}
\left( E_a E_b \right)^{\star} & = &  E_b^{\star} E_a^{\star}~,\\
\left( E_a E_b \right)^{\dag}  & = &  E_b^{\dag} E_a^{\dag}~,
\end{eqnarray*}
$E^{\dag} \equiv E^{\bullet \star} \equiv E^{\star \bullet}$. We shall show 
that the reversion can be used to represent the hermitian 
conjugation.

Let us analyze the following products: 
$\Psi^{\bullet} \Psi$, $\Psi^{\star} \Psi$, $\Psi^{\dag} \Psi$, which involve
the three involutions defined within the Clifford algebra $Cl_{3,0}$.
We must consider the two possibilities due to the identification
of the complex imaginary unit $i=\sqrt{-1}$ by $\sigma_{21}$ and 
$\sigma_{123}$. Let us
perform a real projection of these products,
\begin{eqnarray*}
\left( \Psi^{\bullet} \Psi \right)_S 
 & ~=_{(i\equiv \sigma_{21})}~ & \{ \left[ 
                      \left( \varphi_1 + \sigma_{21} \eta_1 +
                             \sigma_{23} \varphi_2 + \sigma_{13} \eta_2 
                      \right) - \sigma_{123}
                      \left( \varphi_3 + \sigma_{21} \eta_3 +
                             \sigma_{23} \varphi_4 + \sigma_{13} \eta_4
                      \right) \right] \times \\
 & &                  \left[ 
                      \left( \varphi_1 + \sigma_{21} \eta_1 +
                             \sigma_{23} \varphi_2 + \sigma_{13} \eta_2 
                      \right) + \sigma_{123}
                      \left( \varphi_3 + \sigma_{21} \eta_3 +
                             \sigma_{23} \varphi_4 + \sigma_{13} \eta_4
                      \right) \right] \}_S\\
 & &  = \varphi_1^2 - \varphi_2^2 + \varphi_3^2 - \varphi_4^2 -
\eta_1^2    - \eta_2^2    - \eta_3^2    - \eta_4^2~,\\
 & ~=_{(i\equiv \sigma_{123})}~ & \{ \left[ 
                      \left( \varphi_1 + \sigma_{21} \varphi_2 +
                             \sigma_{23} \varphi_3 + \sigma_{13} \varphi_4 
                      \right) - \sigma_{123}
                      \left( \eta_1 + \sigma_{21} \eta_2 +
                             \sigma_{23} \eta_3 + \sigma_{13} \eta_4
                      \right) \right] \times \\
 & &                  \left[ 
                      \left( \varphi_1 + \sigma_{21} \varphi_2 +
                             \sigma_{23} \varphi_3 + \sigma_{13} \varphi_4 
                      \right) + \sigma_{123}
                      \left( \eta_1 + \sigma_{21} \eta_2 +
                             \sigma_{23} \eta_3 + \sigma_{13} \eta_4
                      \right) \right] \}_S\\
 & &  = \varphi_1^2 - \varphi_2^2 - \varphi_3^2 - \varphi_4^2 +
\eta_1^2    - \eta_2^2    - \eta_3^2    - \eta_4^2~,
\end{eqnarray*}
\begin{eqnarray*}
\left( \Psi^{\star} \Psi \right)_S 
 & ~=_{(i\equiv \sigma_{21})}~ & \{ \left[ 
                      \left( \varphi_1 - \sigma_{21} \eta_1 -
                             \sigma_{23} \varphi_2 - \sigma_{13} \eta_2 
                      \right) + \sigma_{123}
                      \left( \varphi_3 - \sigma_{21} \eta_3 -
                             \sigma_{23} \varphi_4 - \sigma_{13} \eta_4
                      \right) \right] \times \\
 & &                  \left[ 
                      \left( \varphi_1 + \sigma_{21} \eta_1 +
                             \sigma_{23} \varphi_2 + \sigma_{13} \eta_2 
                      \right) + \sigma_{123}
                      \left( \varphi_3 + \sigma_{21} \eta_3 +
                             \sigma_{23} \varphi_4 + \sigma_{13} \eta_4
                      \right) \right] \}_S\\
 & &  = \varphi_1^2 + \varphi_2^2 - \varphi_3^2 - \varphi_4^2 +
\eta_1^2    + \eta_2^2    - \eta_3^2    - \eta_4^2~,\\
 & ~=_{(i\equiv \sigma_{123})}~ & \{ \left[ 
                      \left( \varphi_1 - \sigma_{21} \varphi_2 -
                             \sigma_{23} \varphi_3 - \sigma_{13} \varphi_4 
                      \right) + \sigma_{123}
                      \left( \eta_1 - \sigma_{21} \eta_2 -
                             \sigma_{23} \eta_3 - \sigma_{13} \eta_4
                      \right) \right] \times \\
 & &                  \left[ 
                      \left( \varphi_1 + \sigma_{21} \varphi_2 +
                             \sigma_{23} \varphi_3 + \sigma_{13} \varphi_4 
                      \right) + \sigma_{123}
                      \left( \eta_1 + \sigma_{21} \eta_2 +
                             \sigma_{23} \eta_3 + \sigma_{13} \eta_4
                      \right) \right] \}_S\\
 & &  = \varphi_1^2 + \varphi_2^2 + \varphi_3^2 + \varphi_4^2 -
\eta_1^2    - \eta_2^2    - \eta_3^2    - \eta_4^2~,
\end{eqnarray*}
\begin{eqnarray*}
\left( \Psi^{\dag} \Psi \right)_S 
 & ~=_{(i\equiv \sigma_{21})}~ & \{ \left[ 
                      \left( \varphi_1 - \sigma_{21} \eta_1 -
                             \sigma_{23} \varphi_2 - \sigma_{13} \eta_2 
                      \right) - \sigma_{123}
                      \left( \varphi_3 - \sigma_{21} \eta_3 -
                             \sigma_{23} \varphi_4 - \sigma_{13} \eta_4
                      \right) \right] \times \\
 & &                  \left[ 
                      \left( \varphi_1 + \sigma_{21} \eta_1 +
                             \sigma_{23} \varphi_2 + \sigma_{13} \eta_2 
                      \right) + \sigma_{123}
                      \left( \varphi_3 + \sigma_{21} \eta_3 +
                             \sigma_{23} \varphi_4 + \sigma_{13} \eta_4
                      \right) \right] \}_S\\
 & &  = \varphi_1^2 + \varphi_2^2 + \varphi_3^2 + \varphi_4^2 +
\eta_1^2    + \eta_2^2    + \eta_3^2    + \eta_4^2~,\\
 & ~=_{(i\equiv \sigma_{123})}~ & \{ \left[ 
                      \left( \varphi_1 - \sigma_{21} \varphi_2 -
                             \sigma_{23} \varphi_3 - \sigma_{13} \varphi_4 
                      \right) - \sigma_{123}
                      \left( \eta_1 - \sigma_{21} \eta_2 -
                             \sigma_{23} \eta_3 - \sigma_{13} \eta_4
                      \right) \right] \times \\
 & &                  \left[ 
                      \left( \varphi_1 + \sigma_{21} \varphi_2 +
                             \sigma_{23} \varphi_3 + \sigma_{13} \varphi_4 
                      \right) + \sigma_{123}
                      \left( \eta_1 + \sigma_{21} \eta_2 +
                             \sigma_{23} \eta_3 + \sigma_{13} \eta_4
                      \right) \right] \}_S\\
 & &  = \varphi_1^2 + \varphi_2^2 + \varphi_3^2 + \varphi_4^2 +
\eta_1^2    + \eta_2^2    + \eta_3^2    + \eta_4^2~.
\end{eqnarray*}

The first conclusion should be the use of the involution $\dag$ and the 
assumption of a ``real'' geometry. Thus, we should translate 
\[ \left( \, \psi_1^{*} ~ \psi_1^{*} ~ \psi_1^{*} ~ \psi_1^{*} \, \right)
   \left( \begin{array}{c} \psi_1 \\ \psi_2 \\ \psi_3 \\ \psi_4
          \end{array}
   \right) ~ \equiv ~ \sum_{m=1}^4 \, \left( \varphi_m^2+\eta_m^2 \right)
\]
by
\[ 
\left( \Psi^{\dag} \Psi \right)_S~.
\]   
Nevertheless, this real projection of inner products gives an {\em undesired} 
orthogonality between $1$, $\sigma_{21}$ and $\sigma_{123}$. We know that the 
complex imaginary unit, $i=\sqrt{-1}$, represents a phase in the standard 
quantum 
mechanics, thus if we wish to adopt the identifications 
\[ 
i=\sqrt{-1}~~~\rightarrow~~~ \sigma_{21}~~\mbox{or}~~\sigma_{123}~,
\]
we must abandon the ``real'' geometry. We have another possibility. Let us 
rewrite $\Psi$ as follows
\[ \Psi = h_1 + \sigma_{123} h_2 ~~~~~~~ 
h_{1,2} \in {\cal H}(1,\sigma_{21},\sigma_{23},\sigma_{31})~,
\]
the full $\Psi^{\dag} \Psi$ product is given by
\[ \Psi^{\dag} \Psi = \left( h_1^{\dag} - \sigma_{123} h_2^{\dag} \right) 
                      \left( h_1        + \sigma_{123} h_2      \right) =
                      |h_1 |^2 + | h_2 |^2 + 
                   \sigma_{123} \left( h_1^{\dag} h_2 - \mbox{h.c.} \right)~.  
\]  
and so
\[ \Psi^{\dag} \Psi ~=~\mbox{Real Part}~+~\mbox{Vectorial Part}~.\]
Consequently,
\begin{eqnarray*}
\left( \Psi^{\dag} \Psi \right)_S & \equiv & 
\left( \Psi^{\dag} \Psi \right)_{(1,\sigma_{21})}~~~~~~~
~~\sigma_{21} \mbox{-complex geometry}~,\\
\left( \Psi^{\dag} \Psi \right)_S & \equiv & 
\left( \Psi^{\dag} \Psi \right)_{(1,\sigma_{123})}~~~~~~~
~\sigma_{123} \mbox{-complex geometry}~.
\end{eqnarray*}
Now, $(1, \sigma_{21})$ and  $(1,\sigma_{123})$ do not represent orthogonal 
states, and our spinor $\Psi$ have four complex orthogonal states, the
complex orthogonality freedom degrees 
needed to connect a general element of the Pauli 
algebra to the 4-dimensional Dirac spinor
\begin{center}
\begin{tabular}{lll}
$\sigma_{21}$-complex geometry : &
~~~~~$1 \, ,~\sigma_{1} \, , ~\sigma_{23} \, , ~ \sigma_{123}  $ & 
~~~orthogonal states~,\\
$\sigma_{123}$-complex geometry : &
~~~~~$1 \, ,~\sigma_{21} \, , ~\sigma_{23} \, , ~ \sigma_{31} $ & 
~~~orthogonal states~.
\end{tabular}
\end{center}

\section{Barred Operators}

We justify the choice of a complex geometry by noting that although there 
is the possibility to define an anti-self-adjoint operator, $\vec{\partial}$,
with all the properties of a translation operator, {\em imposing a non-complex 
geometry}, there is {\em no} corresponding self-adjoint operator with all the
properties expected for a momentum operator. We can overcome such
a difficulty by using a {\em complex} scalar product and defining as the
{\em appropriate momentum operator}
\begin{center}
\begin{tabular}{ll}
$\sigma_{21}$-complex geometry &
~~~~~$\vec{p} \equiv -\vec{\partial}  \mid \sigma_{21}~, $ \\
$\sigma_{123}$-complex geometry &
~~~~~$\vec{p} \equiv - \sigma_{123} \vec{\partial}~, $ 
\end{tabular}
\end{center}
where $1 \mid \sigma_{21}$ indicates the right action of the bivector 
$\sigma_{21}$. For $\sigma_{123}$, it is not important to distinguish between 
left and right action because $\sigma_{123}$ commutes with all the elements in
$Cl_{3,0}$. Note that the choice 
$\vec{p} \equiv - \sigma_{21} \vec{\partial}$ still gives a self-adjoint 
operator with the standard commutation relations with the coordinates, but
such an operator does not commute with the Hamiltonian, which will, 
in general, be an element of $Cl_{3,0}$. Obviously, in order to write
equations that are relativistically covariant, we must treat the space 
components and time in the same way, hence we are obliged to modify the 
standard ``complex'' equations by the following substitutions
\begin{center}
\begin{tabular}{ll}
$\sigma_{21}$-complex geometry &
~~~~~$i\partial^{\mu} ~ \rightarrow ~ \partial^{\mu}  \mid \sigma_{21}~, $ \\
$\sigma_{123}$-complex geometry &
~~~~~$i\partial^{\mu} ~ \rightarrow ~ \sigma_{123} \partial^{\mu}~. $ 
\end{tabular}
\end{center}

Let us now introduce the complex/linear barred operators. Due to the 
non-commutative nature of the elements of $Cl_{3,0}$, we must distinguish 
between left and right action of $\sigma_{21}$, $\sigma_{23}$, $\sigma_{31}$.
Explicitly, we write
\begin{equation}
\label{ra} 
1\mid \sigma_{21} ~ , ~ 1\mid \sigma_{23} ~ , ~ 1\mid \sigma_{31} ~ , 
\end{equation}
to identify the right multiplication of 
$\sigma_{21}$, $\sigma_{23}$, $\sigma_{31}$, 
\[ 
\left( 1\mid \sigma_{21} \right) \Psi \equiv \Psi \sigma_{21}~,~
\left( 1\mid \sigma_{23} \right) \Psi \equiv \Psi \sigma_{23}~,~
\left( 1\mid \sigma_{31} \right) \Psi \equiv \Psi \sigma_{31}~.
\]
Note that the right action of $\sigma_{1}$, $\sigma_{2}$, $\sigma_{3}$
can be immediately obtained form the operators in (\ref{ra}) by 
$\sigma_{123}$ multiplication.

In rewriting the Dirac equation, we need to work with ``complex'' linear 
barred operators. Here, we must distinguish between $\sigma_{21}$ and 
$\sigma_{123}$ complex geometry. In fact, by working with a 
$\sigma_{123}$-complex 
geometry it is immediate to prove that
\[
1\mid \sigma_{21} ~ , ~ 1\mid \sigma_{23} ~ , ~ 1\mid \sigma_{31} ~ , 
\]
represent $\sigma_{123}$-complex/linear operators. On the contrary, by 
working with
a $\sigma_{21}$-complex geometry we have only one permitted right action, that
is
\[ 1 \mid \sigma_{21}~,\]
which represents a $\sigma_{21}$-complex/linear operator. Why this counting
of parameters? It is simple. In $Cl_{3,0}$ we work with 8 real parameters,
but the most general
linear transformation which can be performed on an element of $Cl_{3,0}$,
adopting a $\sigma_{123}$-complex geometry, is
\[ A + B \mid \sigma_{21} + C \mid \sigma_{23} + D \mid \sigma_{31}~~~~~~~
A,B,C,D \in Cl_{3,0}~,\]
which contains 32 real parameters, the same number of $M_4({\cal C})$. 
This explains the possibility of a direct translation 
between $4\times 4$ complex matrices and 
the Pauli algebra with 
$\sigma_{123}$-complex geometry
\begin{eqnarray*}
\left( \begin{array}{c} \psi_1 \\ \psi_2 \\ \psi_3 \\ \psi_4
       \end{array} \right) & ~~ \leftrightarrow ~~ & 
\Psi =  \psi_1 +
\sigma_{21}           \psi_2 +
\sigma_{23}           \psi_3 +
\sigma_{31}           \psi_4 \\ 
M_4({\cal C})              & ~~ \leftrightarrow ~~ & 
A + B \mid \sigma_{21} + C \mid \sigma_{23} + D \mid \sigma_{31}~.
\end{eqnarray*}

\subsection{$\sigma_{123}$-complex geometry and Dirac equation}

We have now all the tools to reproduce the Dirac equation within the algebra
$Cl_{3,0}$. It is sufficient to translate the standard equation
\[ i \Gamma^{\mu} \partial_{\mu} \Psi = m \Psi~, \]
by using the identification of $i=\sqrt{-1}$ by $\sigma_{123}$ and finding a 
representation of the Dirac matrices, $\Gamma^{\mu}$, 
by elements of the Pauli algebra.  
We observe that the $\Gamma^{\mu}$'s can be rewritten in terms of elements of 
$Cl_{3,0}$, by adopting pseudoscalar and left/right action of bivectors. 
To reproduce 
the right anticommutation relation which characterize the Dirac algebra, 
we perform the following identification
\[ \vec{\Gamma} \sim \left (\sigma_{23}, \sigma_{31}, \sigma_{12} \right)~.\]
To satisfy the anticommutation relation between $\Gamma^0$ and 
$\vec{\Gamma}$, we introduce right actions
\[ \Gamma^0 \sim 1 \mid \sigma_{32} ~~~~~\mbox{and}~~~~~
   \Gamma^{1,2,3} \sim 1  \mid \sigma_{31}~. \]
Finally, the hermiticity conditions give
\begin{eqnarray*}
\Gamma^0 & \equiv & \sigma_{123} \mid \sigma_{32}~,\\
\Gamma^1 & \equiv & \sigma_{123} \sigma_{23} \mid \sigma_{31}~,\\
\Gamma^2 & \equiv & \sigma_{123} \sigma_{31} \mid \sigma_{31}~,\\
\Gamma^3 & \equiv & \sigma_{123} \sigma_{12} \mid \sigma_{31}~.
\end{eqnarray*}
The Dirac equation reads
\begin{equation}
\partial_t \Psi \sigma_{23} + 
\sigma_{23} \partial_x \Psi \sigma_{13} +
\sigma_{31} \partial_y \Psi \sigma_{13} +
\sigma_{12} \partial_x \Psi \sigma_{13} = m \Psi~.
\end{equation}
Let us multiply the previous equation by the barred operator 
$\sigma_{123} \mid \sigma_{23}$,
\[ 
\sigma_{123} \partial_t \Psi \sigma_{23} \sigma_{23} + 
\sigma_{123} \sigma_{23} \partial_x \Psi \sigma_{13} \sigma_{23} +
\sigma_{123} \sigma_{31} \partial_y \Psi \sigma_{13} \sigma_{23} +
\sigma_{123} \sigma_{12} \partial_x \Psi \sigma_{13} = 
m \sigma_{123} \Psi \sigma_{23}~.
\]
By observing that
\[ \sigma_{23}^2 = -1~,~~
\sigma_{13} \sigma_{23}=\sigma_{21}~,~~
\sigma_{123} ~ \left( \sigma_{23}, \sigma_{13}, \sigma_{12} \right)
= - \left( \sigma_{1}, \sigma_{2}, \sigma_{3} \right)~,
\]
we find 
\begin{equation}
\label{de1}
\sigma_{123} \partial_t \Psi + 
\sigma_{1} \partial_x \Psi \sigma_{21} +
\sigma_{2} \partial_y \Psi \sigma_{21} +
\sigma_{3} \partial_x \Psi \sigma_{21} = 
m \Psi \sigma_{1}~,
\end{equation}
which represents the {\em Dirac equation in the Pauli algebra with 
$\sigma_{123}$-complex geometry}. 
This equation is obtained by simple translation, so it reproduces 
the standard physical contents. We are now ready to perform the desired 
translation rules:
\begin{eqnarray*}
\Psi \equiv \left( \begin{array}{c} 
\varphi_1 + i \eta_1 \\
\varphi_1 + i \eta_2 \\
\varphi_1 + i \eta_3 \\
\varphi_1 + i \eta_4 
\end{array} \right) & ~~~\leftrightarrow~~~ &
                      \left( \varphi_1 + \sigma_{123} \eta_1 \right) +
\sigma_{21}           \left( \varphi_2 + \sigma_{123} \eta_2 \right) +
\sigma_{23}           \left( \varphi_3 + \sigma_{123} \eta_3 \right) +
\sigma_{31}           \left( \varphi_4 + \sigma_{123} \eta_4 \right)~,\\
\Phi^{\dag} \Psi ~~~~~~~~~~~     & ~~~\leftrightarrow~~~ & 
\left( \Phi^{\dag} \Psi \right)_{(1,\sigma_{123})}~.  
\end{eqnarray*}
To give the correspondence rules between $4\times 4$ complex matrices and 
barred operators, we need to list only the matrix representations for the
following barred operators 
\[ 1~,~~\sigma_{21}~,~~\sigma_{23}~,~~\sigma_{123}~,
~~1\mid \sigma_{12}~,~~ 1\mid \sigma_{23}~,
\]
all the other operators can be quickly obtained by suitable multiplications 
of the previous ones. The translation of $1$ and $\sigma_{123}$ is very simple:
\[ 1\leftrightarrow \openone_{4\times 4}~~~~~~~\mbox{and}~~~~~~~ 
\sigma_{123} \leftrightarrow i \openone_{4\times 4}~. \] 
The remaining four operators are represented by
\begin{eqnarray*}
\sigma_{21} ~ \leftrightarrow ~ 
\left( \begin{array}{cccc}
0   &  $-1$   &  0      &   0 \\
1   &   0     &  0      &   0 \\
0   &   0     &  0      &   $-1$ \\
0   &   0     &  1   &   0
\end{array} \right) & ~~~~~~~
1 \mid \sigma_{21} ~ \leftrightarrow ~ 
\left( \begin{array}{cccc}
0   &  $-1$   &  0      &   0 \\
1   &   0     &  0      &   0 \\
0   &   0     &  0      &   1 \\
0   &   0     &  $-1$      &   0
\end{array} \right)~,\\ 
\sigma_{23} ~ \leftrightarrow ~ 
\left( \begin{array}{cccc}
0   &   0     &  $-1$   &   0 \\
0   &   0     &  0      &   1 \\
1   &   0     &  0      &   0 \\
0   &   $-1$     &  0      &   0
\end{array} \right) & ~~~~~~~
1 \mid \sigma_{23} ~ \leftrightarrow ~ 
\left( \begin{array}{cccc}
0   &   0        &  $-1$   &   0 \\
0   &   0        &  0      &  $-1$ \\
1   &   0        &  0      &   0 \\
0   &   1     &  0      &   0
\end{array} \right)~. 
\end{eqnarray*}

\subsection{$\sigma_{21}$-complex geometry and Dirac equation}

Let us now discuss the possibility to write down the Dirac equation in the 
Pauli algebra with a $\sigma_{21}$-complex geometry. At first glance a problem
appears. We have not the needed parameters in the barred operators to perform
a translation. In fact, the most general $\sigma_{21}$-complex/linear 
operator is
\[ A + B \mid \sigma_{21}~~~~~~~A,B \in Cl_{3,0}~,\]
and consequently we count only 16 real parameters. We have no hope to 
settle  
down the 32 real parameters characterizing a generic $4\times 4$ 
complex matrix. 
Nevertheless, we must observe the possibility to perform the grade 
involution, {\em which represents a $\sigma_{21}$-complex/linear operation}
\[ \left[ \Psi \left( \alpha + \sigma_{21} \beta \right) \right]^{\bullet} =
   \Psi^{\bullet} 
   \left( \alpha + \sigma_{21} \beta \right)~~~~~~~~~
\alpha, \beta \in {\cal R}~.
\]
Thanks to this involution we double our real parameters. Let us show the 
desired translation rules
\begin{eqnarray*}
\Psi \equiv \left( \begin{array}{c} 
\varphi_1 + i \eta_1 \\
\varphi_1 + i \eta_2 \\
\varphi_1 + i \eta_3 \\
\varphi_1 + i \eta_4 
\end{array} \right) & ~~~\leftrightarrow~~~ &
                      \left( \varphi_1 + \sigma_{21} \eta_1 \right) +
\sigma_{23}           \left( \varphi_2 + \sigma_{21} \eta_2 \right) +
\sigma_{123}          \left( \varphi_3 + \sigma_{21} \eta_3 \right) +
\sigma_{123} \sigma_{23}  \left( \varphi_4 + \sigma_{21} \eta_4 \right)~,\\ 
\Phi^{\dag} \Psi ~~~~~~~~~~~     & ~~~\leftrightarrow~~~ & 
\left( \Phi^{\dag} \Psi \right)_{(1,\sigma_{21})}~.  
\end{eqnarray*}
To give the correspondence rules between $4\times 4$ complex matrices and 
barred operators, we need to list only the matrix representations for the
following barred operators
\[ 1~,~~\sigma_{21}~,~~\sigma_{23}~,~~\sigma_{123}~,
~~1\mid \sigma_{21}~,
\]
and give the matrix version of the grade involution.  
All the other operators can be quickly obtained by 
suitable combinations of the previous operations. 
The translation of $1$ and $1 \mid \sigma_{21}$ is soon obtained:
\[ 1\leftrightarrow \openone_{4\times 4}~~~~~~~\mbox{and}~~~~~~~ 
1 \mid \sigma_{21} \leftrightarrow i \openone_{4\times 4}~. \] 
The remaining rules are
\[
\sigma_{21} ~ \leftrightarrow ~ 
i \, \left( \begin{array}{cccc}
1   &   0     &  0      &   0 \\
0   & $-1$    &  0      &   0 \\
0   &   0     &  1      &   0 \\
0   &   0     &  0      &  $-1$
\end{array} \right)~,~~~ 
\sigma_{23} ~ \leftrightarrow ~ 
\left( \begin{array}{cccc}
0   &   $-1$     &  0      &   0 \\
1   &  0   &  0      &   0 \\
0   &   0     &  0      &  $-1$ \\
0   &   0     &  1      &  0
\end{array} \right)~,~~~ 
\sigma_{123} ~ \leftrightarrow ~ 
\left( \begin{array}{cccc}
0   &   0     &  1      &   0 \\
0   &   0   &  0      &   1 \\
$-1$   &   0     &  0      &   0 \\
0   &  $-1$     &  0      &  0
\end{array} \right)~,
\]
and finally the grade involution is represented by the following matrix
\[ \bullet \mbox{-involution} ~ \leftrightarrow ~
\left( \begin{array}{cccc}
1   &   0     &  0      &   0 \\
0   &   1   &  0      &   0 \\
0   &   0     &  $-1$      &   0 \\
0   &   0     &  0      &  $-1$
\end{array} \right)~.
\]

Let us examine how to translate the Dirac equation
\[ i \Gamma^{\mu} \partial_{\mu} \Psi = m \Psi ~,\]
by working with a $\sigma_{21}$-complex geometry. Firstly, we modify the 
previous equation by multiplying it by $\Gamma^0$ on the left
\[ i \partial_t \Psi + i \Gamma^0 \vec{\Gamma} \cdot \vec{\partial} \Psi
   = m \Gamma^0 \Psi~.\]
We observe that (by using the standard representation~\cite{ZU,BD} for the
Dirac matrices)
\begin{eqnarray*}
\Gamma^0 \Psi & \equiv & \left( \begin{array}{cccc} 
1 & 0 & 0     & 0 \\
0 & 1 & 0     & 0 \\
0 & 0 & $-1$  & 0 \\
0 & 0 & 0     & $-1$ \end{array} \right)
\left( \begin{array}{c} 
\varphi_1 + i \eta_1 \\
\varphi_2 + i \eta_2 \\
\varphi_3 + i \eta_3 \\
\varphi_4 + i \eta_4 
\end{array} \right)\\ 
& ~\leftrightarrow~ &
\left( \varphi_1 + \sigma_{21} \eta_1 \right) +
\sigma_{23}           \left( \varphi_2 + \sigma_{21} \eta_2 \right) -
\sigma_{123}              \left( \varphi_3 + \sigma_{21} \eta_3 \right) -
\sigma_{123} \sigma_{23}  \left( \varphi_4 + \sigma_{21} \eta_4 \right) \\
 & ~\leftrightarrow~&
\left[ 
\left( \varphi_1 + \sigma_{21} \eta_1 \right) +
\sigma_{23}           \left( \varphi_2 + \sigma_{21} \eta_2 \right) +
\sigma_{123}              \left( \varphi_3 + \sigma_{21} \eta_3 \right) +
\sigma_{123} \sigma_{23}  \left( \varphi_4 + \sigma_{21} \eta_4 \right)
                      \right]^{\bullet}~,
\end{eqnarray*}
and
\[ 
\Gamma^0 \vec{\Gamma} ~ \leftrightarrow ~  
\left( \sigma_{1}, \sigma_{2}, \sigma_{3} \right)~,~~~~~
i \openone_{4\times 4} ~\leftrightarrow~ 
1 \mid \sigma_{21}~.
\]                   
Thus, the {\em translated Dirac equation} reads:
\begin{equation}
\partial_t \Psi \sigma_{21} + 
\sigma_{1} \partial_x \Psi \sigma_{21} +   
\sigma_{2} \partial_y \Psi \sigma_{21} +   
\sigma_{3} \partial_z \Psi \sigma_{21} = m \Psi^{\bullet}~.
\end{equation}

\section{Complex Geometries Equivalence}

In the previous sections, we have performed {\em two} 
translated versions of the Dirac equation. Explicitly, 
\begin{equation}
\label{d1}
\sigma_{123} \mbox{-complex geometry}~~~~~~~~~~
\left( \sigma_{123} \partial_t  + \nabla \mid \sigma_{21} \right) \Psi 
= m \Psi \sigma_{1}~,
\end{equation}
and
\begin{equation}
\label{d2}
\sigma_{21} \mbox{-complex geometry}~~~~~~~~~~~~~
\left( \partial_t  + \nabla \right) \Psi \sigma_{21}
= m \Psi^{\bullet}~,
\end{equation}   
where 
\[ \nabla \equiv \sigma_{1} \partial_x +
                 \sigma_{2} \partial_y +   
                 \sigma_{3} \partial_z ~.
\]
We discuss in this section the possibility to relate the two equations 
obtained by imposing different geometries. Let us start by 
taking the $\bullet$-involution of Eq.~(\ref{d1}) 
\begin{equation}
\label{d3}
\sigma_{123} \partial_t \Psi^{\bullet} + \nabla \Psi^{\bullet} \sigma_{21}  
= m \Psi^{\bullet} \sigma_{1}~.
\end{equation}
By working with Eqs.~(\ref{d1},\ref{d3}) we can reobtain Eq.~(\ref{d2}).
To do it, we introduce the idempotents
\[ e_{\pm} = \frac{1}{2} \left( 1 \pm \sigma_{3} \right)~,\]
and give some relations which will be useful in the following 
\[ \left[ e_{\pm}, \, \sigma_{21} \right] = 0~,~~~
\sigma_{1} e_{\pm} = e_{\mp} \sigma_{1}~,
\]
and
\begin{equation}
\label{w} 
\sigma_{123} e_{-} = e_{-} \sigma_{21}~,~~~~~ 
   \sigma_{123} e_{+} = - e_{+} \sigma_{21}~. 
\end{equation}
Let us multiply Eqs.~(\ref{d1}) and (\ref{d3}) from the right respectively by
$e_-$ and $\sigma_{1} e_+$,
\begin{eqnarray*}
\sigma_{123} \partial_t \Psi e_-  + \nabla  \Psi e_- \sigma_{21}  
& = & m \Psi e_+ \sigma_{1}~,\\
\sigma_{123} \partial_t \Psi^{\bullet} \sigma_{1} e_+ - 
\nabla \Psi^{\bullet} \sigma_{1} e_+  \sigma_{21}  
& = & m \Psi^{\bullet} e_+~.
\end{eqnarray*}
By using the relations in Eq.~(\ref{w}), 
we can rewrite the previous equations as follows
\begin{equation}
\label{a} 
\left( \partial_t + \nabla \right) \Psi e_- \sigma_{21} 
= m \Psi e_+ \sigma_{1}~,
\end{equation}
and
\begin{equation}
\label{b}
\left( \partial_t +  \nabla \right)  \Psi^{\bullet} \sigma_{1} e_+ \sigma_{21}
= - m \Psi^{\bullet} e_+~.
\end{equation}
By taking the ``difference'' between these last two equations, we have
\[
\left( \partial_t +  \nabla \right)  
\left[ \Psi e_- - \Psi^{\bullet} \sigma_{1} e_+ \right] \sigma_{21}
= m \left[ \Psi e_+ \sigma_{1} + \Psi^{\bullet} e_+ \right]~.
\]
By redefining
\begin{equation}
\Phi \equiv \Psi e_- - \Psi^{\bullet} \sigma_{1} e_+ ~,
\end{equation}
and noting that
\[
\Phi^{\bullet} =   \Psi^{\bullet} e_+  + \Psi \sigma_{1} e_-        
                       =   \Psi^{\bullet} e_+  + \Psi e_+ \sigma_{1}~,
\]
we find
\begin{equation}
\left( \partial_t + \nabla \right) \Phi  \sigma_{21} 
= m \Phi^{\bullet}~,        
\end{equation} 
as anticipated. 

We conclude this section by discussing the phase transformations 
characterizing our equations. 
It is immediate to show that the phase 
transformation             
\[ \Psi \rightarrow \Psi e^{\sigma_{123} \alpha}~~~~~~~ 
\alpha \in {\cal R}~,\]
implies the following transformation on $\Phi$ 
\[ \Phi \rightarrow \Phi e^{\sigma_{21} \alpha}~. \]
In fact, 
\begin{eqnarray*}
\Phi' & = & \Psi e^{\sigma_{123} \alpha} e_- - 
                    \Psi^{\bullet} e^{-\sigma_{123} \alpha} \sigma_{1} e_+ \\
              & = & \Psi e_- e^{\sigma_{21} \alpha} - 
                    \Psi^{\bullet}  \sigma_{1} e_+ e^{\sigma_{21} \alpha}\\
              & = & \Phi e^{\sigma_{21} \alpha}~.
\end{eqnarray*}
At this stage, there is not difference in the using a $\sigma_{123}$ or 
$\sigma_{21}$ complex geometry. So, we have an 
equivalence between $\sigma_{123}$ and $\sigma_{21}$ complex geometry
{\em within the Pauli algebra}. 

\section{Conclusion}

The possibility of using Clifford algebra to describe standard quantum
mechanics receives a major thrust with the adoption of a complex scalar 
product (complex geometry). A second important step in this objective of 
translation is achieved with the introduction of the so-called barred 
operators, which permit to write down few translation rules which allow to 
quickly reproduce in the $Cl_{3,0}$ formalism the standard results of the 
Dirac theory. All the relations can be manipulated without introducing a 
matrix representation, greatly simplifying the algebra involved.

In this paper we worked with the Pauli algebra but we wish to remark that 
our considerations can be immediately generalized to the spacetime algebra, 
which represents the natural language for relativistic quantum mechanics.

In the standard literature, the unit scalar imaginary of quantum mechanics is
replaced by a bivector. We showed that another possibility is also available, 
namely the identification of the unit scalar imaginary $i=\sqrt{-1}$ by the
pseudoscalar $\gamma_{0123}$ of the spacetime algebra ($\sigma_{123}$ in the
Pauli algebra). These two geometric interpretations reflect the {\em two} 
possible choices in defining a complex geometry within the multivector
formalism. At the free-particle level, there is an equivalence in using these 
two complex scalar products.

We conclude by observing that a possible difference between the 
$\sigma_{21}$ and
$\sigma_{123}$ complex geometries could appear in the formulation of the 
Salam-Weinberg model, where the electromagnetic group is obtained by symmetry
breaking from the Glashow group $SU(2) \times U(1)$. It appears natural
to use
\[ \sigma_{21}~,~~~\sigma_{23}~,~~~\sigma_{31}~~~~~\mbox{and}~~~~~
   1 \mid \sigma_{21}~,
\]
as generators of the electroweak group. In this case the right choice
should be the adoption of a $\sigma_{21}$ complex geometry. After symmetry
breaking the remaining electromagnetic group will be identified by the
left/right action of the generator $\sigma_{21}$. A complete discussion of the
Salam-Weinberg model within the multivector formalism will be given in a
forthcoming paper~\cite{DRV}.

\acknowledgements

One of the authors (SdL) enjoyed the help of many friends and colleagues:
Cap\'{\i}, Dermevalle, Marcelo, Angela, Evelize, Liliam, Paula e Vera .  
In particular, the author would like to thank Mario e Dora for the   
hospitality during the stay in Brasil, and Luis for his genuine friendship. 
For financial support, SdL is indebted to the IMECC-UNICAMP.


\end{document}